\documentclass[]{aa}  

\usepackage{graphicx}
\usepackage{txfonts}
\usepackage[]{hyperref}
\usepackage{amsmath}
\usepackage{amssymb}
\usepackage{textcomp}
\usepackage{xcolor}

\newcommand{\ct}{$^{13}{\rm C}$}
\newcommand{\nf}{$^{14}{\rm N}$}
\newcommand{\nfi}{$^{15}{\rm N}$}
\newcommand{\ms}{$M_{\odot}$~}
\newcommand{\cd}{$^{12}{\rm C}$}
\newcommand{\fl}{$^{19}{\rm F}$}
\newcommand{\nqa}{${\rm ^{15}N(\alpha, \gamma)^{19}F}$}

\begin{document} 

\title{Magnetic-buoyancy-induced mixing in AGB Stars: fluorine nucleosynthesis at different metallicities}

\author{D. Vescovi \inst{1}\fnmsep\inst{2}\fnmsep\inst{3}
\and S. Cristallo\inst{3}\fnmsep\inst{2}
\and S. Palmerini\inst{4}\fnmsep\inst{2}
\and C. Abia\inst{5}
\and M. Busso\inst{4}\fnmsep\inst{2}
}

\institute{Goethe University Frankfurt, Max-von-Laue-Strasse 1, Frankfurt am Main 60438, Germany\\
\email{vescovi@iap.uni-frankfurt.de}
\and INFN, Section of Perugia, Via A. Pascoli snc, 06123 Perugia, Italy
\and INAF, Observatory of Abruzzo, Via Mentore Maggini snc, 64100 Teramo, Italy
\and Department of Physics and Geology, University of Perugia, via A. Pascoli snc, 06123 Perugia, Italy
\and University of Granada, Departamento de Fisica Teorica y del Cosmos, 18071 Granada, Spain
}

\date{Received ; accepted }

 

\abstract{
Asymptotic giant branch (AGB) stars are considered to be among the most significant contributors to the fluorine  budget in our Galaxy.
While at close-to-solar metallicity observations and theory agree, at lower metallicities stellar models overestimate the fluorine production with respect to heavy elements.
We present \fl~nucleosynthesis results for a set of AGB models with different masses and metallicities in which magnetic buoyancy acts as the driving process for the formation of the \ct~neutron source (the so-called \ct~pocket).
We find that \fl~is mainly produced as a result of nucleosynthesis involving secondary \nf~during convective thermal pulses, with a negligible contribution from the \nf~present in the \ct~pocket region.  
A large \fl~production is thus prevented, resulting in lower fluorine surface abundances. As a consequence, AGB stellar models with magnetic-buoyancy-induced mixing at the base of the convective envelope well agree with available fluorine spectroscopic measurements at both low and close-to-solar metallicity.}

\keywords{Stars: abundances -- Stars: AGB and post-AGB -- Stars: magnetic field -- Magnetohydrodynamics (MHD) -- Stars: carbon -- Nuclear reactions, nucleosynthesis, abundances}

\maketitle
%

\section{Introduction}\label{sec:intro}
Understanding the cosmic origin of fluorine is one of the most interesting topics in the nuclear astrophysics field.
The production of its sole stable isotope, \fl, largely depends from the physical conditions characterizing stellar interiors. To date, it is widely debated which is the primary source of fluorine in the Universe and several sites have been proposed as potential candidates:
AGB stars \citep{forestini92}, rapidly rotating massive stars \citep{limongi18}, Wolf–Rayet stars \citep{meynet00}, core-collapse supernovae \citep{woosley88}, and novae \citep{jose98}.
While the contributions from many of these sources are required to explain the galactic chemical evolution of fluorine abundance (see e.g. \citealt{meynet00,renda04,kobayashi11,spitoni18,prantzos18,olive19,grisoni20}), the only direct observation of fluorine production is provided by spectroscopic findings of photospheric [F/Fe] enhancements in intrinsic AGB carbon stars \citep{jorissen92,abia09,abia10,abia11,abia19} and metal-poor extrinsic stars \citep{lucatello11,abia19}.

AGB stars consist of a degenerate C-O core, surrounded by a He-shell and a H-shell, separated by a He-rich intermediate zone (He-intershell), and an extended convective envelope. 
Recurrently, on a timescale of tens of thousand years of quiescent H-shell burning, the He-burning shell becomes thermally unstable.
The large amount of energy released during this thermal pulse (TP), mainly due to the triple-$\alpha$ reaction, induces convective motions throughout the He-intershell and causes an expansion, and thus a cooling of the H-burning shell. As a consequence, the latter is switched off until the star contracts, and the rising temperature is large enough to its re-ignition.
This process repeats up to the complete erosion of the envelope by stellar winds and characterizes the thermally pulsing AGB (TP-AGB) phase of these stars (e.g., \citealt{iben83}).
During the expansion of the envelope, convection may penetrate deep into the H-He discontinuity beyond the H-burning shell and carry to the surface fresh products of the nucleosynthesis.
This phenomenon is known as third dredge-up (TDU).
At each TDU episode, protons are partially mixed from the envelope into the He-intershell and, as the H-burning shell reignites, are captured by the abundant \cd~nuclei to produce a \ct-enriched layer, the so-called \ct~pocket. 
${}^{13}$C nuclei are then efficiently burned via the ${\rm {}^{13}C(\alpha, n){}^{16}O}$ reaction at $T \approx 9 \times 10^{7} ~\mathrm{K}$ (see \citealt{cristallo18} and references therein), so producing the free neutrons necessary for the synthesis of heavy elements in low mass $\mathrm{AGB}$ stars ($M \lesssim 3 M_{\odot}$) through the so-called s(low)-process (see \citealt{busso99,herwig05,straniero06,karakas14} for reviews).
A second neutron burst is driven by the ${\rm {}^{22}Ne(\alpha, n){}^{25}Mg}$, which is efficiently activated at the base of the convective zone produced by a TP when the temperature reaches about $3 \times 10^{8} ~\mathrm{K}$ (mostly in more massive AGBs, see e.g. \citealt{karakas14}).

During the interpulse phase, part of the neutrons released within the He-intershell are captured by ${\rm ^{14}N}$ through the ${\rm ^{14}N(n, p) ^{14}C}$ reaction to synthesize \nfi~by means of the chain ${\rm ^{14}C(\alpha, \gamma) ^{18}O(p, \alpha) ^{15}N}$.
\nfi~is then burnt to primary \fl~via~\nqa~reaction in the subsequent convective TP \citep{goriely00,lugaro04,cristallo14}.
An additional contribution to fluorine production comes from secondary \ct~and \nf~synthesized in the H-burning ashes and from the eventual unburnt \ct~in the pocket \citep{cristallo09}. In both cases, \ct~is engulfed in the convective shell generated by the TP and produce further secondary \fl~through the same above mentioned reactions chain. 
Fluorine is then carried to the surface by convective motions during the TDU. Therefore, its envelope abundance is expected to be correlated with those of carbon and $s$-process elements (see \citealt{abia19} and references therein).

Actually, one of the most debated topics in AGB modeling concerns the formation of the \ct~pocket. The solution to this problem is strictly connected to a deep understanding of the physical processes governing mass-exchange at the interface between the convective envelope and the radiative core.
Latest research has focused on studying non-convective transport mechanisms, usually ignored by the canonical theory of stellar structure and evolution.
Different kinds of additional transport processes were invoked for the penetration of proton-rich material from the convective envelope into the He-intershell: convective overshoot \citep{herwig97}, opacity-induced overshoot \citep{cristallo09,cristallo11,cristallo15b}, mixing induced by rotation \citep{herwig03,siess04,piersanti13} or internal gravity waves \citep{denissenkov03,battino16}, and magnetic-buoyancy-induced mixing \citep{trippella14,trippella16}.
Focusing on the production of fluorine, it was suggested that a mechanism leading to the formation of an extended \ct~pocket and, at the same time, to a small amount of \nf, could solve the problem of \fl~overproduction with respect to s-elements in low-mass metal-poor objects \citep{abia19}. 

Post-process neutron-capture models for AGB stars, in which the formation of the required \ct~reservoir is ascribed to magnetic-buoyancy-induced mixing, were shown to be able to account for the solar distributions of \textit{s}-only isotopes \citep{trippella16}, isotopic ratios of \textit{s}-elements measured in presolar SiC grains \citep{palmerini18} and a large part of abundance observations in evolved low-mass stars \citep{busso21}. All the aforementioned works applied the theory of magnetic-induced buoyancy mixing theorized by \cite{nucci14}. This theory has recently been adopted by \citet{vescovi20}, who contextualized magnetic mixing in the more complex framework of FRUITY stellar evolutionary models, obtaining a satisfactory fit to presolar grain isotopic ratios. The same stellar models also provide a consistent explanation for the observed yttrium abundance trends in the Galactic disk's inner region \citep{magrini21}.
FRUITY Magnetic models predict that deep profiles of low proton abundances are generated below the convective envelope border. 
The low proton concentration severely limits local \nf~formation, as protons are almost entirely consumed for the synthesis of \ct, potentially affecting \fl~production as well.

Here we investigate fluorine nucleosynthesis in low-mass AGB stars by computing a new series of stellar models accounting for the formation of a magnetic-buoyancy-induced \ct~pocket.
The structure of the paper is as follows.
In Section~\ref{sec:models} we present the stellar evolutionary code, the nuclear network, and the mixing algorithm adopted to calculate the AGB models. 
In Section~\ref{sec:results} we compare our models with available F spectroscopic abundances in a sample of Galactic and extra-Galactic AGB carbon stars, as well as with other metal-poor Ba-type and CH-type evolved stars. Finally, in the last section we summarize our conclusions.

\section{Stellar models}\label{sec:models}
The stellar models presented in this work have been computed using the FUNS code (see \citealt{straniero06} and references therein) by following the chemical evolution of approximately 500 isotopes (from hydrogen to bismuth) linked by more than 800 reactions (charged particle reactions, neutron captures, and $\beta$-decays).
The baseline nuclear network is essentially the same already described in \citet{cristallo09} with the addition of some recently updated reaction rates. In particular, the list of adopted reaction rates relevant to the nucleosynthesis of fluorine is reported in Table~\ref{tab:input_rates}.
\begin{table}[!t]
\caption{Reaction rates of relevance to fluorine nucleosynthesis used in our computations.}
\label{tab:input_rates}
\centering
\begin{tabular}{cc}
\hline \hline
Reaction & Reference \\
\hline
${\rm {}^{14}N(p,\gamma){}^{15}O}$ & 1 \\
${\rm {}^{15}N(p,\gamma){}^{16}O}$ & 2 \\
${\rm {}^{17}O(p,\gamma){}^{18}F}$ & 3 \\
${\rm {}^{18}O(p,\gamma){}^{19}O}$ & 4 \\
${\rm {}^{15}N(p,\alpha){}^{12}C}$ & 5 \\
${\rm {}^{17}O(p,\alpha){}^{14}N}$ & 6 \\
${\rm {}^{18}O(p,\alpha){}^{15}N}$ & 7 \\
${\rm {}^{19}F(p,\alpha){}^{15}O}$ & 8 \\
${\rm {}^{14}C(\alpha,\gamma){}^{18}O}$ & 9 \\
${\rm {}^{14}N(\alpha,\gamma){}^{18}F}$ & 10 \\
${\rm {}^{15}N(\alpha,\gamma){}^{19}F}$ & 10 \\
${\rm {}^{18}O(\alpha,\gamma){}^{22}Ne}$ & 10 \\
${\rm {}^{19}F(\alpha,p){}^{22}Ne}$ & 11 \\
${\rm {}^{13}C(\alpha,n){}^{16}O}$ & 12 \\
${\rm {}^{14}N(n, p){}^{14}C}$ & 13\\ 
\hline
\end{tabular}
 \tablebib{
 (1)~\citet{imbriani05}; (2) \citet{leblanc10}; (3) \citet{dileva14}; (4) \citet{best19};
 (5) \citet{angulo99}; (6) \citet{bruno16}; (7) \citet{bruno19}; (8) \citet{indelicato17};
 (9) \citet{johnson09}; (10) \citet{iliadis10}; (11) \citet{dagata18}; (12) \citet{trippella17}; \citet{wallner16}.
 }
\end{table}

For radiative opacities, we computed opacity tables by means of the OPAL Web tool\footnote{\url{https://opalopacity.llnl.gov/new.html}}.
At low temperatures $(\log T \leq 4.05$), typical of the external layers of stars, we used the ÆSOPUS tool \citep{marigo09}, which includes molecular and atomic species relevant for AGB atmospheres, to compute the opacity tables.
In particular, ÆSOPUS allows to take into account changes in opacity due to both eventual oxygen and $\alpha$-enhancement (see Table~\ref{tab:input_models}) and to chemical composition variations of the envelope when the star becomes carbon-rich \citep{marigo02,cristallo07,ventura10}.
\begin{table}[!t]
\caption{
Initial mass, [Fe/H], total metallicity (comprehensive of $\alpha_{\rm O-Ca}$ enhancements), and helium adopted for the models presented in this work
}
\label{tab:input_models}
\centering
\begin{tabular}{cccccc}
\hline \hline
Mass ($M_\odot$)& [Fe/H] & Z & Y & $\alpha_{\rm O}$ & $\alpha_{\rm Ne-Ca}$ \\
\hline
1.5 & $-2.18$ & $3.10 \times 10^{-4}$ & 0.247 & 0.7 & 0.4 \\
1.5 & $-1.88$ & $5.33 \times 10^{-4}$ & 0.248 & 0.6 & 0.4 \\
1.5 & $-1.70$ & $8.00 \times 10^{-4}$ & 0.248 & 0.6 & 0.4 \\
1.5 & $-1.18$ & $2.13 \times 10^{-3}$ & 0.250 & 0.5 & 0.4 \\
2.0 & $-0.70$ & $5.70 \times 10^{-3}$ & 0.254 & 0.4 & 0.3 \\
2.0 & $-0.40$ & $8.13 \times 10^{-3}$ & 0.257 & 0.2 & 0.15 \\
2.0 & $-0.18$ & $1.00 \times 10^{-2}$ & 0.259 & 0 & 0 \\
2.0 & $+0.05$ & $1.67 \times 10^{-2}$ & 0.267 & 0 & 0 \\
2.0 & $+0.13$ & $2.00 \times 10^{-2}$ & 0.271 & 0 & 0 \\
\hline
\end{tabular}
\end{table}
Opacity tables used in this work have been calculated with the scaled-solar element distribution by \citet{lodd20}. Accordingly, a solar-calibrated value of the mixing length parameter $\alpha_{\rm ml} = 1.86$ is adopted (see \citealt{vescovi19} for more details).
Regarding the mass-loss rate, we adopted a Reimers' formula with $\eta = 0.4$ for the pre-AGB evolution, and the rate used by \citet{abia20} for the AGB phase (such a rate is slightly different with respect to \citealt{straniero06}, due to the adoption of recalculated bolometric corrections).

The surface enrichment of C and s-process elements in AGB stars is related to the complex coupling between convective mixing and nuclear processes.
The problem of the neutron source in AGB stars, and in particular the physical process driving the formation of a \ct~pocket in the He-rich intershell is still a matter of debate (see Section~\ref{sec:intro}).
In past FRUITY models, the partial mixing of hydrogen from the envelope necessary to produce fresh \ct~was accounted for by the so-called opacity-induced overshoot \citep{straniero06,cristallo09}.
This powers an extra-mixing and a formation of a chemically smooth transition zone between the fully convective envelope and the radiative region.
At the inner border of the convective envelope, the velocity of the descending material accelerated by convection is estimated as
\begin{equation}
v = v_{\rm cb} \, \text{exp} \left(-\frac{\delta r}{\beta H_P} \right),
\end{equation}
where $v_{\rm cb}$ is the velocity at the convective border, $\delta r$ is the distance from the border, $H_P$ is the pressure scale height at the convective border, and $\beta$ is a free parameter whose value was set to 0.1 in order to maximize the $s$-process production in low-mass AGB stars (see \citealt{cristallo09,guan13}).
The free parameter $\beta$ regulates the amount of protons mixed beyond the bare convective border, and also affects the TDU efficiency. 
In standard FRUITY models, the partial mixed zone is extended below the formal Schwarzschild convective boundary down to $2~H_P$ \citep{straniero06}. Instead, in FRUITY Tail models \citep{cristallo15a} the penetration limit is fixed to the layer where the convective velocity is $10^{-11}$ times lower than the value attained at the Schwarzschild border. The ensuing \ct~pockets were shown to be remarkably larger than those obtained in standard FRUITY models, thus resulting in a substantial increase of surface s-process enrichment.\\
\citet{vescovi20} found that, due the inclusion of updated physical inputs, the slope of the exponential decline of convective velocities has to be reduced down to $\beta = 0.025$ not to alter the efficiency of the TDU and obtain a sizable amount of dredged-up material. 
However, models computed with such parameter choice show an extra-mixing so inefficient to almost inhibit the production of $s$-process nuclei. Thus, mixing triggered by magnetic fields has been introduced as an additional mixing mechanism. 
As in \cite{nucci14}, these authors assumed that a toroidal field is present in the radiative He-intershell at the beginning of the TDU and triggers the buoyant rise of magnetic flux tubes. 
As a consequence, a matter flow is pushed to the envelope. This induces, for mass conservation, a matter down-flow of H-rich material to He-intershell necessary for the formation of the \ct~pocket. 
The down-flow velocity can be expressed as 
\begin{equation}\label{eq:vfinal1}
v(r) = u_{\rm p}  \left( \frac{r_{\rm p}}{r} \right)^{k+2},
\end{equation}
where $r_{\rm p}$ is the distance from the stellar center from which magnetic flux tubes, generated in the He-intershell, start to rise. $r_{\rm p}$ can be identified from the critical toroidal $B_{\varphi}$ necessary for the onset of magnetic buoyancy instabilities \citep{vescovi20}. 
$k$ is the exponent quantifying the density decline, being $\rho(r) \propto r^{k}$, in the He-rich radiative layers below the convective envelope during a TDU and it is typically lower than -3 (see also \citealt{busso21}).
The identification of the critical field necessary for the occurrence of instabilities by magnetic buoyancy allows to identify the corresponding radial position $r_{\rm p}$ from which magnetic structures arise.
$u_{\rm p}$ is the \emph{effective} starting buoyant velocity.
Actually, the magnetized domains are rather fast and occupy only a small fraction of the total mass of the stellar layer ($\simeq 1/10^5$, see e.g. \citealt{busso07,trippella16}), thus transporting mass at a rate equal to assuming an \emph{effective} buoyant velocity for all the material in the layer.
In \citet{vescovi20}, it was found that most of the heavy-element isotope ratios measured in presolar SiC grains from AGB stars are consistently explained by stellar AGB models computed with a unique choice for the toroidal field strength and the initial buoyant velocity, namely $B_{\varphi} = 5 \times 10^{4}$ G and $u_{\rm p} = 5 \times 10^{-5}$ cm s$^{-1}$.
Nonetheless, we expect a non-trivial mass-metallicity dependence of the mixing triggered by magnetic buoyancy (as for other type of mixing, see e.g. \citealt{joyce18,battino21}). In particular, the toroidal field strength is determined by the differential rotation profile established in the He-intershell and which amplifies the magnetic field via dynamo action (see Section \ref{conclu}). In this framework, the compactness of a star at the pre-AGB and AGB stages is a fundamental property that influences the dynamo process and the following magnetic-buoyancy-induced mixing. The stellar models presented here have H-exhausted core masses at the beginning of the TP-AGB phase very similar to the $2~M_{\odot}$ close-to-solar metallicity models presented in \citet{vescovi20}. Thus, we adopt for all the computed models their parameter choice for $B_{\varphi}$ and $u_{\rm p}$ without further adjustments. We demand to a future work the analysis of the dependence of the magnetic mixing from the stellar core mass.


%
\section{Results and discussion}\label{sec:results}
\begin{figure}
\centering
\resizebox{\hsize}{!}{\includegraphics{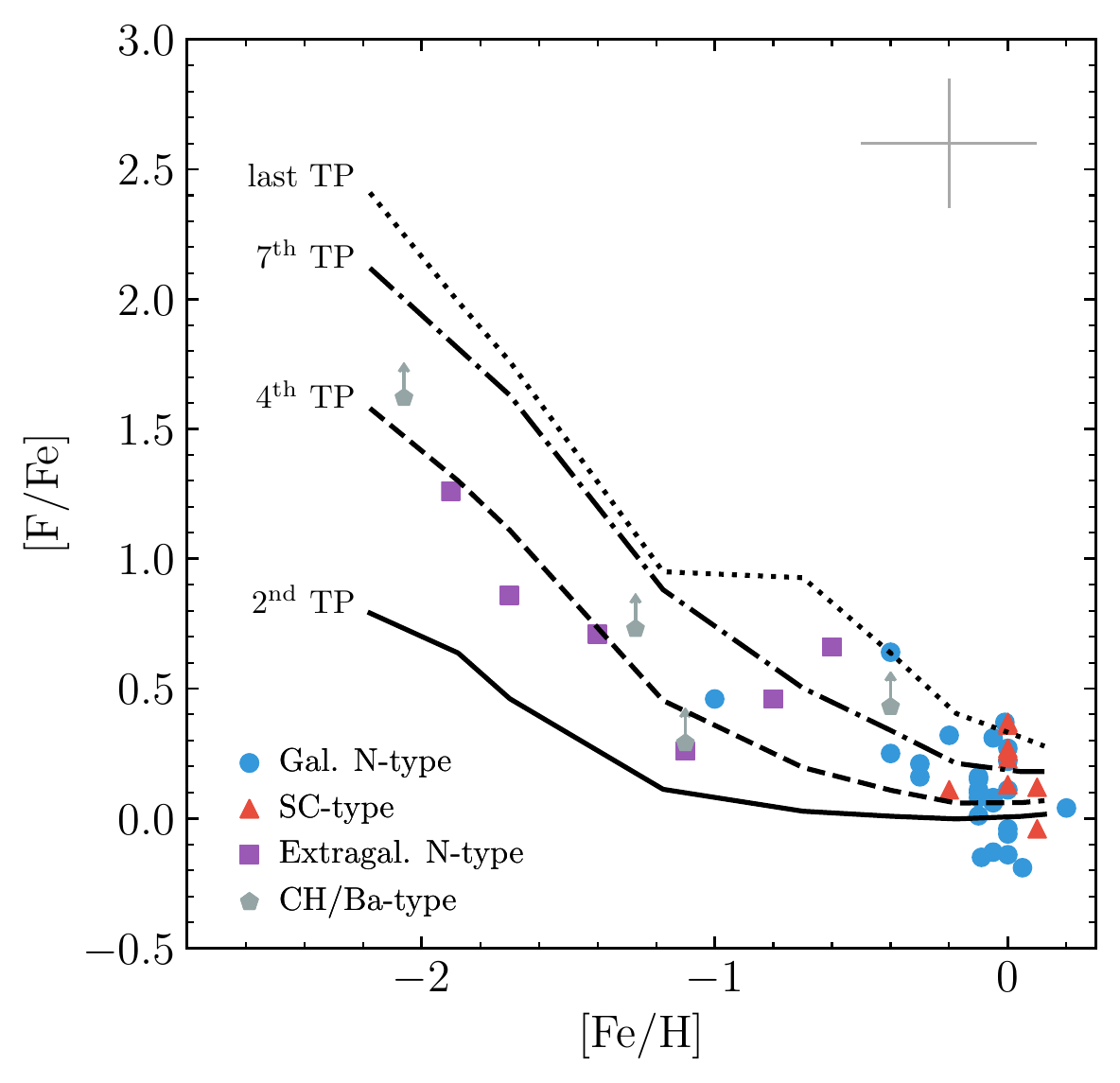}}
\caption{Comparison between observed [F/Fe] ratios as a function of the metallicity and FRUITY Magnetic models. Symbols refer to the four data groups: circles, galactic carbon stars; triangles, SC-type stars; squares, extragalactic carbon stars; pentagons, extrinsic CH/Ba stars. Lines represent theoretical predictions for 2 (for [Fe/H] $ \geq -0.7$) and 1.5 \ms (for [Fe/H] < $-0.7$) AGB stars at different TPs. A typical error bar is indicated.}
\label{fig:fluoro_fe}
\end{figure}
In the following, we compare extant literature data and new FRUITY Magnetic models.
In doing so, we adopt data from previous studies for intrinsic AGB carbon stars \citep{abia10,abia11,abia15,abia19} and extrinsic CH/Ba stars \citep{lucatello11}. 
In Fig.~\ref{fig:fluoro_fe} we report the [F/Fe] ratios of our selected sample as a function of the iron content [Fe/H]. We plot four group data: galactic (N-type) carbon stars, SC-type stars, extragalactic carbon stars, and extrinsic CH/Ba stars. 
Observational data are compared with stellar models of 1.5~\ms and 2~\ms at different TPs. Within the observational errors there is a good agreement, confirming the expected F-enhancement trend with the metallicity (see e.g. \citealt{lugaro04,cristallo14}).
One word of caution has to be used regarding fluorine abundances in extrinsic CH/Ba stars, for which [F/Fe] ratios only represent lower limits due to dilution effects of binary mass-transfer phenomena \citep{abia19}.

\begin{figure}[!t]
\centering
\resizebox{\hsize}{!}{\includegraphics{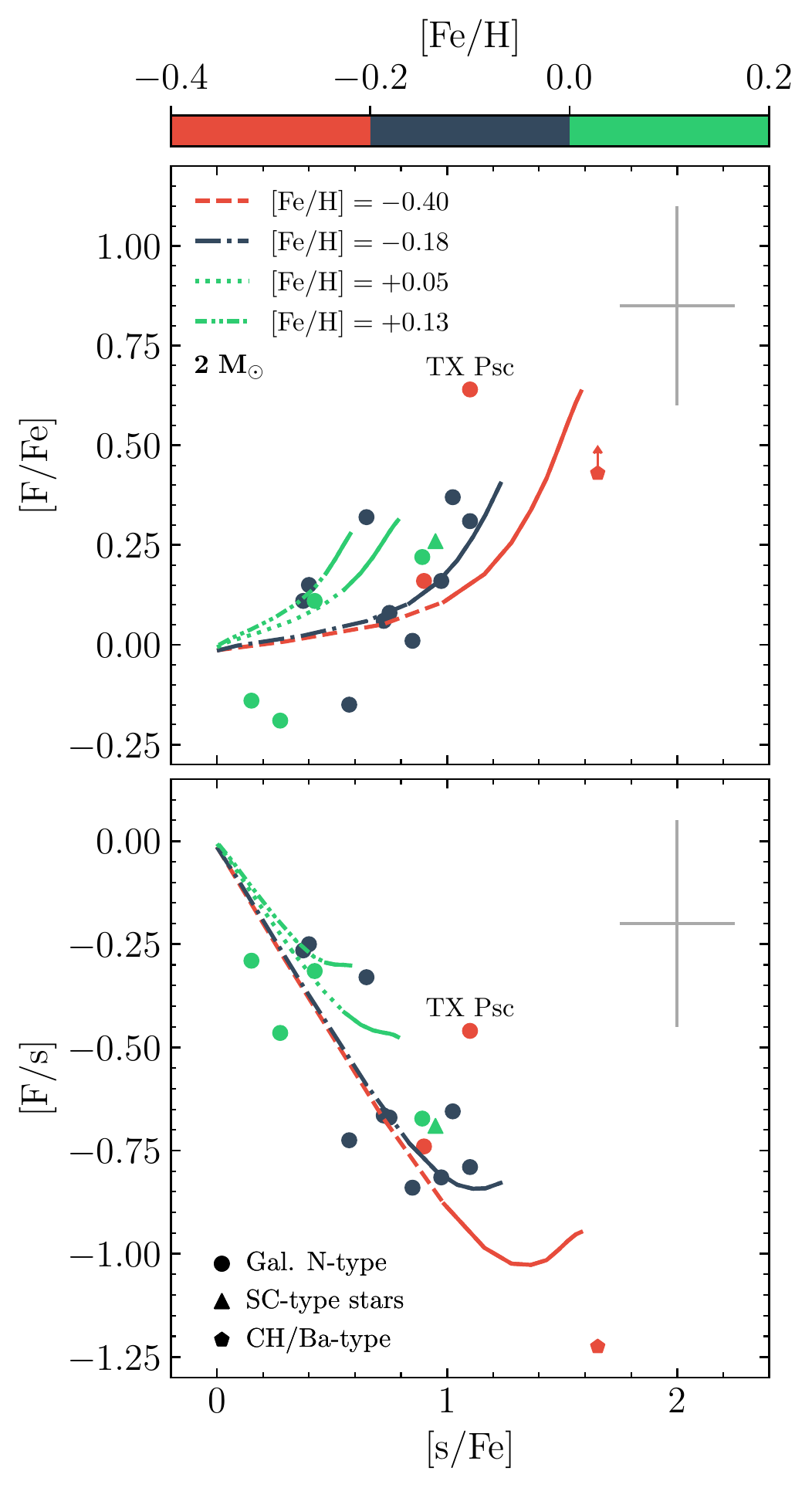}}
\caption{Observed [F/Fe] (top panel) and [F/s] (bottom panel) vs. average s-element enhancements compared with theoretical predictions for 2 \ms and close-to-solar metallicity models. Symbols as in Fig.~\ref{fig:fluoro_fe}. Data points and theoretical lines are color-coded by [Fe/H]. The continuous portion of the lines represents the theoretical C-rich phase, while the discontinuous portion represents the O-rich phase. Typical uncertainties are shown. See text for details.}
\label{fig:fluoro_s}
\end{figure}
The top panel of Fig.~\ref{fig:fluoro_s} shows the comparison between theoretical predictions of FRUITY Magnetic models with initial mass $ M = 2~M_{\odot}$ and [Fe/H] $\geq - 0.4$ and spectroscopic observations for [F/Fe] ratios vs. the average s-element enhancement.
Usually, four observational indices are used to represent s-process distributions and overabundances: [ls/Fe] (representative of the first s-process peak), [hs/Fe] (representative of the second s-process peak), [s/Fe], and [hs/ls].
Given the inhomogeneity of the available elemental abundance of heavy s-elements for the selected sample, we use the mean of the relative abundances of only Y and Zr to express the ls-index, and Ba and La to express the hs-index. The difference between the hs- and the ls-index quantifies the s-element index [hs/ls], while the total average s-element enhancement is given by the mean of the relative abundances of Y, Zr, Ba, and La.
For close-to-solar metallicity galactic N-type carbon stars (dots) and SC-type stars (triangles) the observed increase of fluorine with s-element enhancement is well reproduced by theoretical models. Note that the predicted [F/Fe] are always positive since \fl~is overall produced in low-mass AGB stars (see also \citealt{lugaro04}). On the other hand, a few observational data show negative values which are, however, still consistent, within the errors, with no fluorine production. 
The comparison with N-type and extrinsic CH/Ba type stars (pentagon) with [Fe/H] $< - 0.2$ is more challenging due to the smaller number of available information at these metallicities. In particular, two out of the three stars seems to exhibits the same trend observed at higher metallicity and are in good agreement with theoretical predictions, while the fluorine and s-elements abundances derived for TX Psc is slightly out of the trend and it is not fully reproduced by stellar models.
This interpretation is confirmed even by considering [F/s] ratios (see bottom panel of Fig.~\ref{fig:fluoro_s}). Studying such an index, the fluorine enhancement for the extrinsic stars is not affected by uncertainties related to the dilution factor and provides a more robust tool for comparison. Confirming the previous analysis, models are able to replicate the quasi-linear decreasing trend of [F/s] with the surface s-process enrichment, aside from the marginally-outside TX Psc. Given the typical uncertainties affecting observational data, however, it is difficult to firmly conclude that model predictions and observations are in disagreement for this single star.

\begin{figure}
\centering
\resizebox{\hsize}{!}{\includegraphics{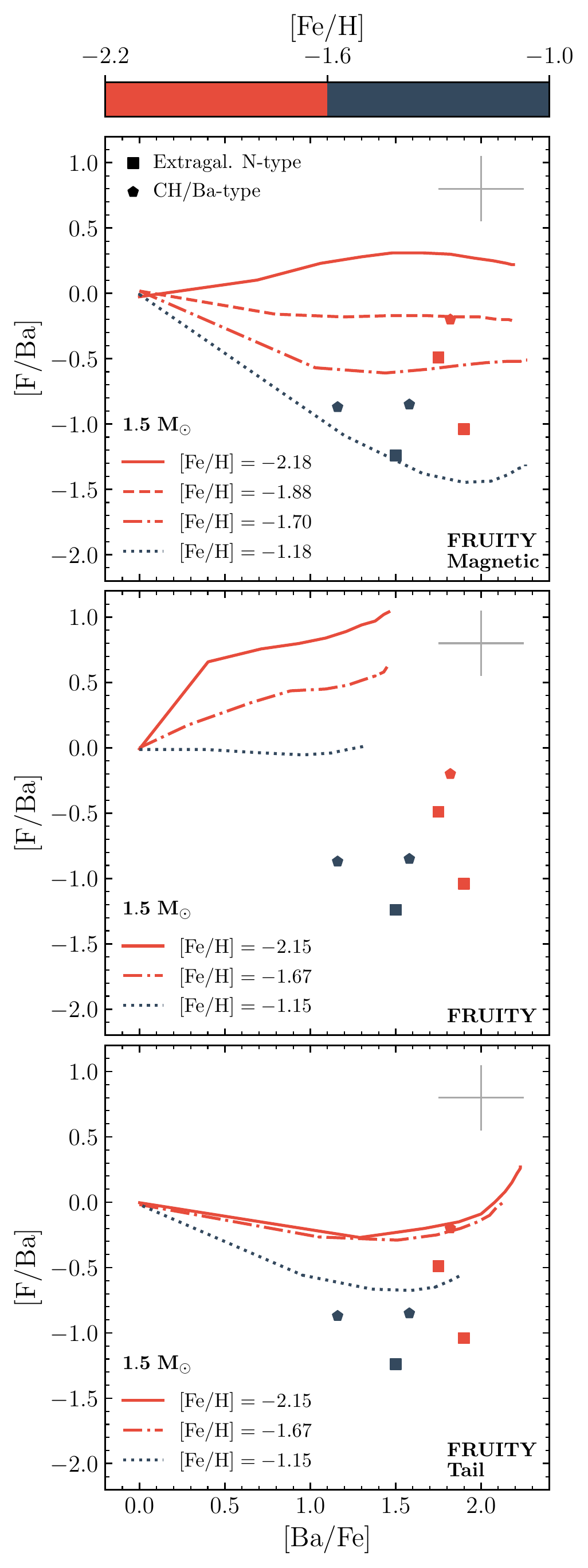}}
\caption{[F/Ba] vs. [Ba/Fe] in the sample stars with [Fe/H] $\leq -1.0$. Symbols: squares, extragalactic carbon stars; pentagons, extrinsic CH/Ba stars.
Lines are theoretical predictions for 1.5 \ms TP-AGB stars with low metallicity, assuming different mechanisms for the formation of the \ct~pocket (see text for details). Note that at these low metallicities, theoretical AGB models predict that the star becomes C-rich from first TDU episodes. Data points and theoretical lines are color-coded by [Fe/H]. Typical error bars are indicated.}
\label{fig:fluoro_ba}
\end{figure}
Note that, in our computations we used the ${\rm {}^{13}C(\alpha,n){}^{16}O}$ reaction rate provided by \citet{trippella17}, which combined the latest asymptotic normalization coefficient values and the Trojan horse method, to determine the astrophysical $S$-factor with a indirect approach. 
At the typical temperature of $\sim 90$ MK corresponding to the radiative \ct~burning phase, the most recent direct measurement, given by \citet{heil08}, is almost $20\%$ higher. Adopting such a rate has however no impact in the production of s-nuclei if the \ct~pocket is fully consumed during the interpulse period (see also \citealt{trippella17}). 
Instead, if some amount of \ct~survives, due to the lower reaction rate, it will be swallowed by the convective shell generated by the following TP and burnt at a rather high temperature ($\sim 200$ MK), so providing an additional neutron burst and possibly affecting the production of s-nuclei and \fl. This burst was shown to occur only during the very first TPs in low-mass AGB stars with high metallicity ($Z \geq 0.01$; see \citealt{cristallo09,karakas10}). In our $2~M_\odot~Z = 0.02$ model, switching from \citet{trippella17} to \citet{heil08} reaction rate produces variations of less than 3\% in [F/Fe], [ls/Fe], and [hs/Fe] indexes, thus  much lower that the typical uncertainties affecting spectroscopic observations of s-process-rich stars. Stellar models of lower metallicity are expected to produce even minor variations \citep{cristallo09,karakas10} unless they have mass and metallicity so low ($M \lesssim 1.3 ~M_{\odot}$ and [Fe/H] $\lesssim -2.5$) to experience a proton ingestion episode at the first fully developed TP (see \citealt{cristallo09b,campbell10,choplin21}).

\begin{figure}
\centering
\resizebox{\hsize}{!}{\includegraphics{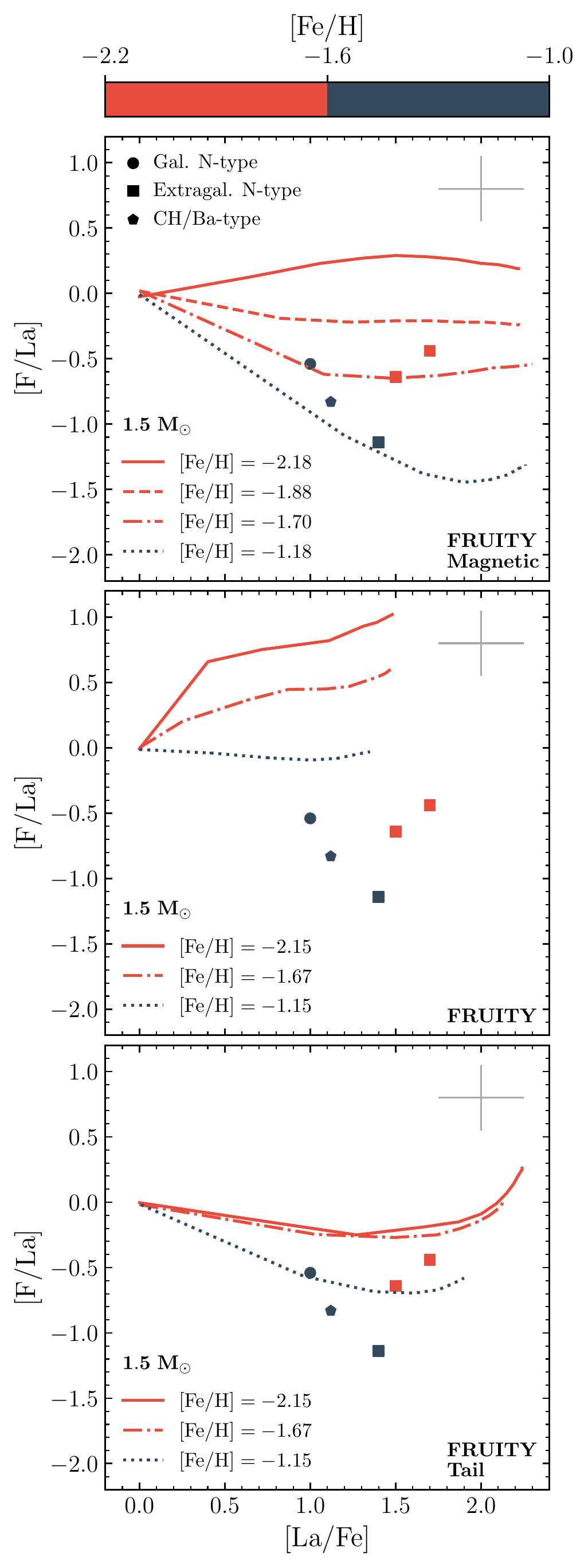}}
\caption{Same as Fig.~\ref{fig:fluoro_ba}, but for [F/La] vs. [La/Fe].}
\label{fig:fluoro_la}
\end{figure}
In Figures~\ref{fig:fluoro_ba}~and~\ref{fig:fluoro_la} we perform a similar comparison at low metallicities. 
Given the fact that there is no homogeneous sample of stars with both Ba and La, we compared model predictions separately. For both figures, we present in different panels theoretical expectations from new FRUITY Magnetic models (top panels), from standard FRUITY models (middle panels) and from FRUITY Tail models (bottom panels). Magnetic models are capable of well reproducing the spread observed, at different metallicities, for both [F/Ba] and [F/La] ratios as a function of the corresponding overall s-process enhancement.
Conversely, standard FRUITY models fail in reproducing both the Ba and La enrichment, as well as the [F/Ba] and [F/La] ratios. Finally, FRUITY Tail models are sufficiently enriched in s-elements, due to the larger \ct~pockets than those obtained in standard FRUITY models, but still show a systematic overproduction of fluorine with respect to Ba and La.
As a whole, FRUITY Magnetic models show a reduction of fluorine production in good agreement to spectroscopic observations for low metallicity stars. 

As previously mentioned, \fl~is primarily synthesized in AGB stars via the \nqa~reaction reaction in the convective zone
generated by a TP. The production of \nfi~in the He-intershell involves a complex nuclear chain of successive n-- (or p--) and $\alpha$--captures starting from \ct~nuclei.
A first source of \ct~is represented by the \ct~pocket itself, whose radiative burning leads to the accumulation of \nfi. A second source is the \ct~left in the H-burning shell ashes, which is engulfed into the convective zone generated by the TP and rapidly burned via the ${\rm {}^{13}C(\alpha, n){}^{16}O}$ reaction. 
In standard FRUITY models, the two channels almost equally contribute to the fluorine nucleosynthesis, especially at close-to-solar metallicity. On the other hand, moving to low metallicities the contribution of \ct~left in the H-burning ashes is less important \citep{cristallo09}.
Instead, in FRUITY Tail models, the larger \ct~pockets guarantee a net increase of s-process abundances and a large fluorine decrease at fixed s-process surface enhancement \citep{abia15}. However, these models barely reach negative values for [F/Ba] and [F/La] (see lower panels of Figures~\ref{fig:fluoro_ba}~and~\ref{fig:fluoro_la}), pointing out that fluorine production need to be further suppressed, without altering the s-process enrichment (see also \citealt{abia19}).
The extended profile and the low proton abundance characterizing FRUITY Magnetic models \citep{vescovi20} have the twofold effect of forming large \ct~pockets and reducing the formation of primary \nf. 
In this framework, the few available protons make first \ct~through the ${\rm ^{12}C(p, \gamma)^{13}C}$ reaction, preventing more proton captures to form \nf~and thereby inhibiting the nuclear chain ${\rm ^{14}N(n, p) ^{14}C(\alpha, \gamma) ^{18}O(p, \alpha) ^{15}N (\alpha, \gamma)^{19}F}$.
Hence, any fluorine appearing in AGB envelopes in these models is of secondary nature, generated by \nf~concentrations left behind by H-shell burning.
This results in low fluorine enhancements and high s-enhancements, that pose FRUITY Magnetic models in close agreement with observations in very metal-poor AGB stars.

\section{Conclusions}\label{conclu}
The fluorine production in low-mass AGB stars has been revisited in the light of upgraded FRUITY stellar models, in which the \ct~neutron source is related to magnetic-buoyancy-induced phenomena.
Predicted fluorine enhancements are in agreement with those observed in carbon stars at different metallicities.
The observed correlation between F and the s-element enhancements is also reproduced.
On one hand, new FRUITY Magnetic models show a reduced net \fl~production, with respect to models in which the partial mixing of hydrogen during a TDU is attributed to overshooting below the convective envelope of an AGB star. This is a consequence of the low abundance of \nf~in the \ct~pocket, which leads to a negligible production of fluorine during the \ct~radiative burning. The \fl~ envelope abundance is therefore ascribed only to the amount of the secondary \ct~ in the H-shell ashes, which depends on the CNO abundances in the star.
On the other hand, mixing induced by magnetic buoyancy leads to extended \ct~pockets so resulting in large surface s-process enrichments. 
As a whole, new FRUITY Magnetic models simultaneously accounts for both the observed fluorine and the average s-element enhancements in intrinsic AGB carbon stars and extrinsic CH/Ba stars.\\
Different types of data (presolar grain measurements and spectroscopic observations) related to stars belonging to different components of our Galaxy (disk and halo) point to a single configuration of mixing induced by magnetic buoyancy. 

The above scenario suggests that magnetism should also be a quite common phenomenon in evolved low-mass stars. In this respect, there exists several observations of magnetic fields of a few G in the envelopes of AGB stars (see, e.g., \citealt{vlemmings19} and references therein). Another piece of evidence for magnetic fields in low-mass stars is provided by the increasing number of observed magnetic white dwarfs (WDs).
During the least decades, many of WDs lying within 20 pc from the Sun revealed to have field strength above $\sim 1$ MG, while for most of them the available data were not sufficient to asses the presence or not of magnetic fields. These studies suggested that about 10\% of all white dwarfs are magnetic (e.g. \citealt{kawka07}). 
However, the advent of increasingly precise spectroscopic and spectropolarimetric observations, which improve the capability to detect magnetic fields of low strength, revealed that at least $\sim 20\%$ of WDs possess magnetic fields ranging from a few kG up to about 1000 MG \citep{landstreet19,bagnulo20}, indicating that magnetism is a widespread, rather than a rare, property of WDs.
These magnetic fields are most likely leftovers of previous phases, and they may be the result of a complicated interaction between fossil and dynamo generated fields during stellar evolution and/or stellar merger events \citep{ferrario20}.
In our scenario, we propose that, below the extended convective envelope of low-mass AGB stars, strong toroidal magnetic fields exist and trigger a magnetic-buoyant-induced mixing adequate for the formation of the \ct~pocket. The required mixing rate put constraints on the field strength.
In this picture, the toroidal field is assumed to originate from a dynamo operating in the AGB interiors which amplifies a small seed poloidal field by draining the available differential rotation energy.

The present study provides another, independent, confirmation (see \citealt{magrini21}) that the calibration performed by \citet{vescovi20} on the free parameters characterizing magnetic buoyancy-induced mixing is extremely robust. However, additional analyses are needed to validate such a statement, before claiming for a definitive theory. 

\begin{acknowledgements}
DV and SC acknowledge S. Bagnulo for
fruitful discussions.
DV acknowledges financial support from the German-Israeli Foundation (GIF No. I-1500-303.7/2019). CA acknowledges financial support from the Agencia Estatal de Investigación of the Spanish Ministerio de Ciencia e Innovación through the FEDER founds projects PGC2018-095317-B-C2.
\end{acknowledgements}

%
\bibliographystyle{aa} 
\bibliography{biblio} 
%

\end{document}